# NOVEL O(H(N)+N/H(N)) ALGORITHMIC TECHNIQUES FOR SEVERAL TYPES OF QUERIES AND UPDATES ON ROOTED TREES AND LISTS

Mugurel Ionuţ ANDREICA

Computer Science Department, Politehnica University of Bucharest, Bucharest, Romania
email: mugurel.andreica@cs.pub.ro

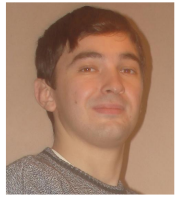

**Ph.D. Eng.
Mugurel Ionuţ ANDREICA**

**Abstract:** *In this paper we present novel algorithmic techniques with a O(H(N)+N/H(N)) time complexity for performing several types of queries and updates on general rooted trees, binary search trees and lists of size N. For rooted trees we introduce a new compressed super-node tree representation which can be used for efficiently addressing a wide range of applications. For binary search trees we discuss the idea of globally rebuilding the entire tree in a fully balanced manner whenever the height of the tree exceeds the value of a conveniently chosen function of the number of tree nodes. In the end of the paper we introduce the H-list data structure which supports concatenation, split and several types of queries. Note that when choosing H(N)=sqrt(N) we obtain O(H(N)+N/H(N))=O(sqrt(N)).*

## 1. INTRODUCTION

Tree-structured networks occur in many fields, like computer networks and distributed systems (where tree topologies are the most efficient solutions for some types of communication) [7], business management (where a company is organized in hierarchical structures or uses some hierarchical approaches to risk management) [12, 13] or data mining and management (where tree-like or list-like data structures are used for organizing the data and performing efficient queries and updates).

In this paper we present novel techniques for solving several types of query and update problems on rooted trees and lists. In Section 2 we present a general technique for constructing a small compressed super-node tree representation of a general rooted tree. Then, using the compressed super-node tree, we can solve many problems more efficiently than the naive straight-forward solutions applied to the whole tree (see Section 3 for some examples). In Section 4 we present a very simple solution for maintaining a binary search tree balanced (for a loose balacing criterion) under an arbitrary sequence of insertions and deletions. Although this solution is not as efficient as many other techniques proposed in the literature, it is very easy to implement. In Section 5 we introduce the H-list data structure which supports concatenation, split and several types of queries. In Section 6 we discuss related work and in Section 7 we conclude and present future work.

## 2. THE COMPRESSED SUPER-NODE TREE REPRESENTATION OF A ROOTED TREE

We present a general technique for constructing a compressed super-node tree representation of a rooted tree with *N* nodes having the following properties (where *H* is a parameter – generally, a function of the number of nodes *N*):

- The nodes of the compressed super-node tree are a subset of *O(N/H)* nodes of the original rooted tree (and the root of the original tree is also the root of the compressed super-node tree).
- The "edge" from a child to its parent in the compressed super-node tree corresponds to a path with *O(H)* nodes from the original tree.
- The "edges" from any two children to their parents in the compressed super-node tree correspond to internally disjoint paths from the original tree.
- Each node of the original tree has one of the following attributes:
    o is a node of the compressed super-node tree
    o is a node on a path corresponding to an edge from a child to its parent in the compressed super-node tree
    o has an ancestor which is either a node of the compressed super-node tree or on a path corresponding to an edge from the compressed super-node tree and this ancestor is located *O(H)* levels above (levels are counted in the original tree)

We will first describe how to construct such a tree in *O(N)* time and then we will show how it can be used as an efficient representation for solving several types of problems on trees.

We will first assign a level to each node of the (original) tree. The level of the root is *level(root)=0* and the level of any other node *x* is *level(x)=level(parent(x))+1* (where *parent(x)* denotes the parent of the node *x* in the tree). We will initially mark as "special" all the nodes *x* whose level is divisible by *H* (i.e. *level(x) mod H = 0)*. Note that the root is marked as special, since its level is *0*. Fig. 1 shows an example of computing the special nodes on a sample tree.

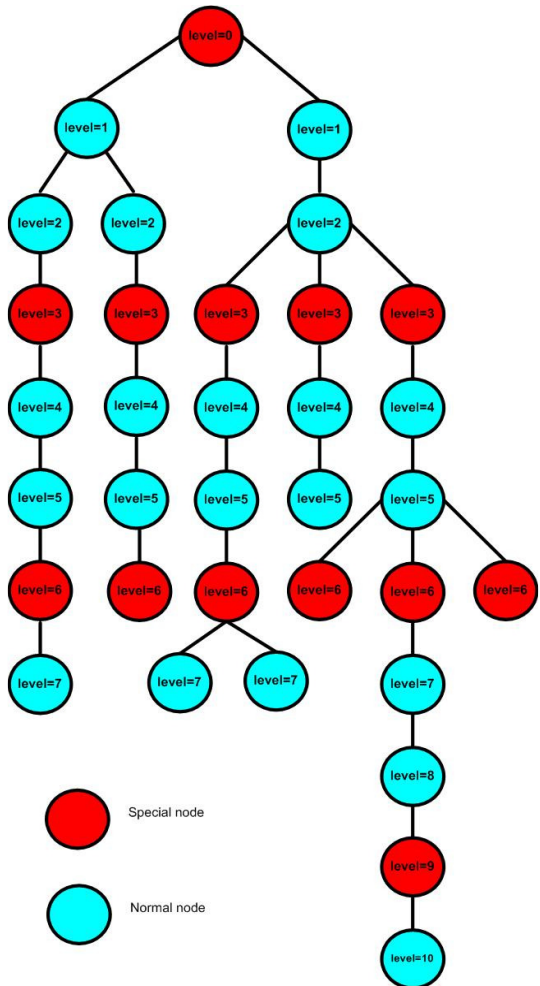

**Fig. 1. Computing the special nodes on a sample tree.**

Then, we visit all the special nodes and for each such node $x \neq root$ we climb up the tree (using the *parent* pointers) until we reach the first special node above it. Let this node be $y$. We will mark $y$ as being a *super-node* (initially, only the root is marked as a super-node). After considering all the special nodes, the super-nodes are those nodes which have at least one special node below them (plus the root of the tree). Let's notice that each super-node has at least $H$ tree nodes below it, because we know that there is at least one special node located below it. Moreover, all the internal nodes on the path from the super-node to the special node below it are not special nodes. Thus, for each super-node $x$ there are $O(H)$ non-special nodes below it and not below any other super-node $y$ located below $x$ (if any). Thus, we will "assign" these $O(H)$ nodes to $x$. By a simple counting argument, there can be at most one super-node for every $O(H)$ non-special nodes, meaning that the number of super-nodes is $O(N/H)$. The super-nodes will be nodes of the compressed super-node tree.

Let's assign to each super-node (except the root) its initial super-parent as the first super-node above it. We may compute these initial super-parents independently for each non-root super-node $x$ by starting from $x$ and following the parent pointers of the original tree until we reach the first super-node $y$ above $x$. Then $y$ will be the initial super-parent of $x$. Since there are $O(N/H)$ super-nodes and the initial super-parent of each super-node is $H$ levels above it, the time complexity of this step is $O(N)$. Fig. 2 shows the (initial) super-nodes obtained for the same tree depicted in Fig. 1.

At this point the problem we have to solve is that the "edge" from super-nodes to their initial super-parents do not cover necessarily internally disjoint paths in the original tree. If two nodes $x$ and $y$ have the same initial super-parent $z$, the path from $x$ to $z$ and the one from $y$ to $z$ may intersect before reaching the node $z$. In order to solve this problem, we will add extra super-nodes. Let the current set of super-nodes be called *ISN* (initial super-nodes) and let *ESN* denote the set of extra super-nodes (initially, *ESN* is empty). We will attempt to visit all the nodes above the super-nodes from *ISN* and select extra super-nodes (if needed) in the process. Before starting, we will assume that none of the original tree nodes was visited.

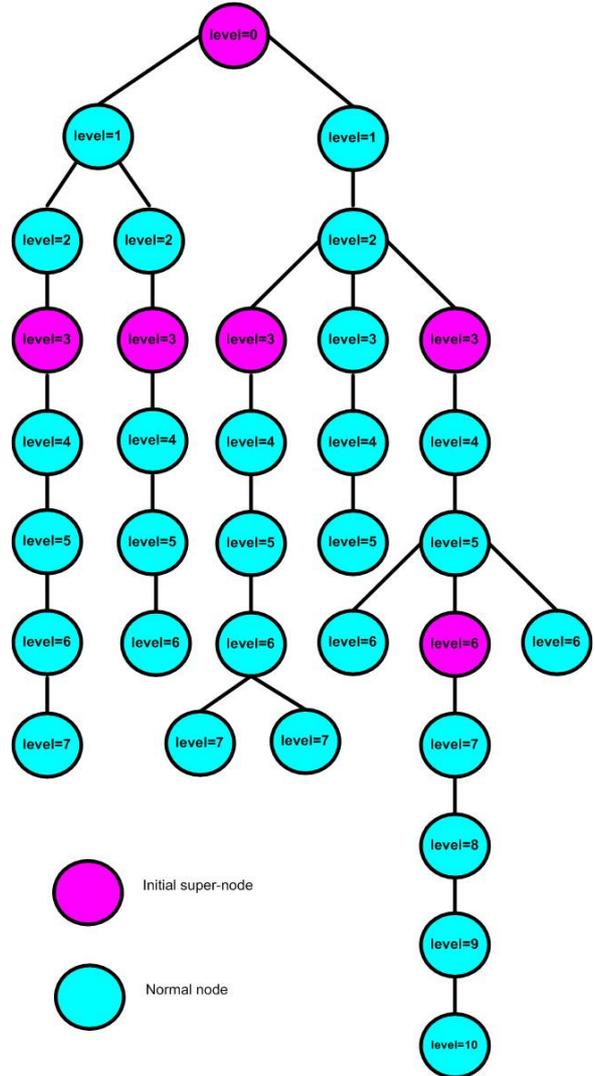

**Fig. 2. (Initial) Super-nodes for the same tree as in Fig. 1.**

We will consider each node $x \neq root$ from *ISN* independently. We will start from $y=parent(x)$. While *(y is not a super-node)* and *(y was not visited)* we: *(1)* mark $y$ as *visited*; *(2)* set $y=parent(y)$. When the while loop ends the node $y$ has one of the following two properties:

- $y$ is a super-node or
- $y$ was marked as visited previously

If $y$ was previously marked as visited (and, thus, $y$ is not a super-node), we will mark $y$ as being a super-node and add it to the set *ESN*. The reason behind this is that, before making $y$ a super-node, $y$ was an intersection node of two or more paths from some initial super-nodes to their common initial super-parent.

Let's notice that, in the worst case, *ESN* may not contain more nodes than *ISN* (each node $x$ from *ISN* adds at most one node *y* to *ESN*). Thus, at the end of this procedure, the number of super-nodes at most doubles (but is still of the order $O(N/H)$).

Now we are ready to compute the final super parents (also called super parents, without the *final* attribute). For each super-node $x \neq root$ we follow the parent pointers starting from *x* until we reach the first super-node *y* located above *x*. *y* will be the (final) super parent of *x* (*super-parent(x)=y*). As before, this step takes only $O(N)$ time overall.

In the end, let's notice that the path from each super-node to its super-parent is internally disjoint from the path of any other super-node to its super-parent (the paths of two super-nodes with the same super-parent will have in common the super-parent, but no other internal nodes). We will associate all the nodes on the tree path from a super-node *x* to *super-parent(x)* (including *x* and excluding *super-parent(x)*) to node *x* (and we will say that they form node *x*'s group). The path from a super-node *x* to *super-parent(x)* is an edge of the compressed super-node tree (and we will also call it a super-edge) and consists of $O(H)$ tree nodes.

Let's analyze now the situation of the tree nodes which are neither super-nodes nor do they belong to a super-edge. Let's consider such a node *z*. *z* has a super-node ancestor located at most $2 \cdot H$ levels above it. Note that the initial selection of special nodes made sure that each node had a special node at most *H* levels above it. Since super-nodes are the special nodes without their lowest layer, the previous condition is updated to $2 \cdot H$ levels above.

This concludes the construction of the compressed super-node tree. All the properties we mentioned in the beginning of this section were satisfied. Note that if we choose $H=O(sqrt(N))$ then the compressed super-node tree has $O(sqrt(N))$ nodes, each group has $O(sqrt(N))$ nodes and each node which is neither a super-node nor a node belonging to a super-node's group has a super-node (or a node belonging to a super-node's group) $O(sqrt(N))$ levels above it. We denote by *sqrt(N)* the square root of *N*. Fig. 3 shows the final result of our algorithm when applied on the same tree as in Fig. 1 and Fig. 2. We present applications of this compressed representation of a rooted tree in the following section.

## 3. APPLICATIONS OF THE COMPRESSED SUPER-NODE TREE REPRESENTATION OF A ROOTED TREE

### 3.1. Extending a block partitioning framework [1] to compressed super-node trees

In [1] a block partitioning framework was introduced for handling a wide range of point and range queries and point and range updates on a sequence of cells (i.e. cells which are arranged sequentially, as if forming a path). A point query asks for the value of a given cell and a point update changes the value of a single cell. A range query asks for the aggregate value of an interval of cells and a range update performs the same operation on each cell of a given interval.

The block partitioning framework from [1] has at its core the following observation. Each query or update interval *[i,j]* can be split into three sub-intervals (whenever *i* and *j* are not in the same block): the interval *[i,a]* consisting of all the cells starting at *i* and ending at the last cell in *i*'s block, the interval *[b,j]* consisting of all the cells ending at *j* and starting at the first cell in *j*'s block, and the interval *[a+1,b-1]* which covers a number of *full blocks*. If *i* and *j* are in the same block then all the cells between *i* and *j* are queries/updated individually. Note that each block contains *H* consecutive cells (the last block may contain fewer); thus, *N* cells can be divided into approximately *N/H* blocks. A range query or update visits $O(H)$ individual cells and $O(N/H)$ full blocks. By maintaining query and update aggregates for each block, each range query and update can be performed in $O(H+N/H)$ time (assuming that the query and update aggregates can be represented with $O(1)$ size). Point queries and updates can be performed in $O(1)$ time.

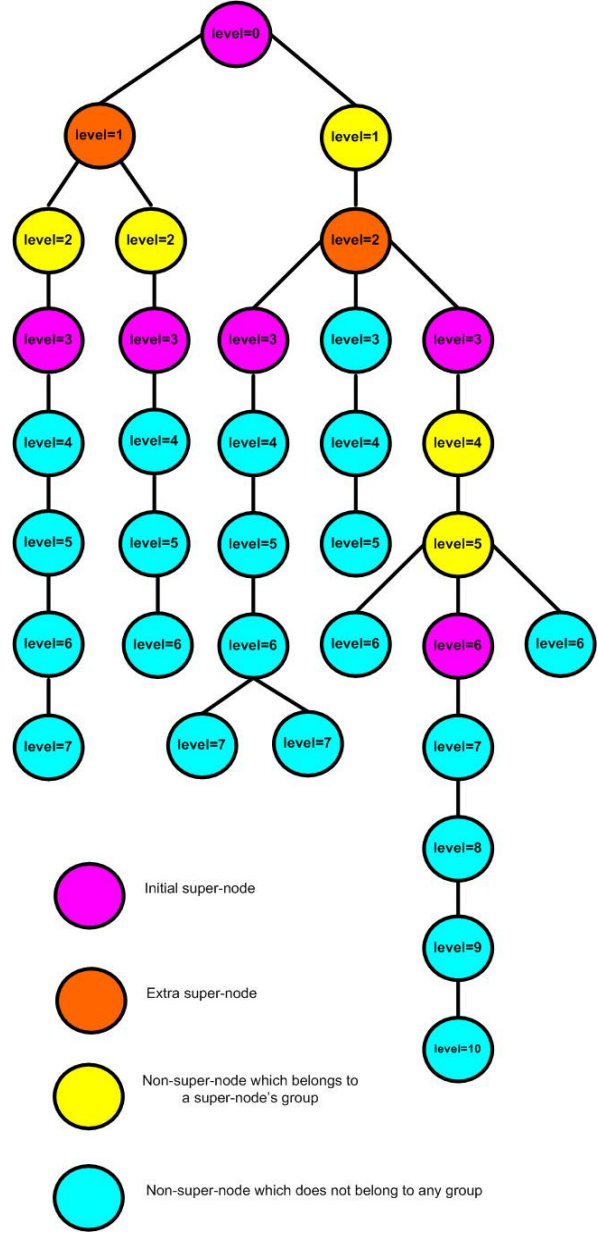

**Fig. 3. Final results of the compressed super-node tree construction algorithm applied on the same tree as in Fig. 1 and Fig. 2.**

Before discussing other applications, we will first extend the applicability of the block partitioning framework introduced in [1] to trees. A block will correspond to a super-node's group. A range query and update will correspond to a query or update on a path in the tree. A query/update path between two nodes *i* and *j* will be handled by first finding the lowest common ancestor *LCA(i,j)*. *LCA(i,j)* can be found using alternative techniques (e.g. [8]), but the compressed super-node tree representation can also be used directly, as shown below:

**FindLCA(i,j):**
**while** *(i≠j)* {
 **if** (*i* and *j* are both super-nodes) **then** {
  **if** (*level(i)≥level(j)*) **then** {
   *i=super-parent(i)*
  } **else** {
   *j=super-parent(j)*
  }
 } **else if** (neither *i* nor *j* is a super-node) **then** {
  **if** (*level(i)≥level(j)*) **then** {
   *i=parent(i)*
  } **else** {
   *j=parent(j)*
  }
 } **else if** (*i* is a super-node and *j* is not a super-node) **then** {
  **if** (*j* belongs to *i*'s group) **then** {
   *i=parent(i)*
  } **else if** (*i≠root* and *level(super-parent(i))≥level(j)*) **then** {
   *i=super-parent(i)*
  } **else** {
   *j=parent(j)*
  }
 } **else** {
  **if** (*i* belongs to *j*'s group) **then** {
   *j=parent(j)*
  } **else if** (*j≠root* and *level(super-parent(j))≥level(i)*) **then** {
   *j=super-parent(j)*
  } **else** {
   *i=parent(i)*
  }
 }
}
**return** *i*

If we maintain group information during the construction of the compressed super-node tree (i.e. to which group each node belongs or the fact that it belongs to no group), then the procedure presented above takes $O(H+N/H)$ time for finding *LCA(i,j)*. Once the LCA is found, a range query/update on the path between a node *i* and the LCA can be performed as follows:

**RangeQueryOrUpdate(i,LCA):**
**while** *(i≠LCA)* {
 **if** (*i* is a super-node) **then** {
  **if** (*level(super-parent(i))≥level(LCA)*) **then** {
   *query or update the full block corresponding to i's group*
   *i=super-parent(i)*
  } **else** {
   *query or update the node i, taking care to consider the fact that it belongs to node i's group*
   *i=parent(i)*
  }
 } **else** {
  **if** (*i* belongs to node *g*'s group) **then** {
   *query or update the node i, taking care to consider the fact that it belongs to node g's group*
   *i=parent(i)*
  } **else** {
   *query or update the node i individually, as in the naive solution (which would simply visit every node on the path); node i does not belong to any group*
   *i=parent(i)*
  }
 }
}

A range query/update on the path between *i* and *j* needs to visit the paths from *i* to *LCA(i,j)* and from *j* to *LCA(i,j)* (by calling *RangeQueryOrUpdate(i,LCA(i,j))* and *RangeQueryOrUpdate(j,LCA(i,j))*). Then, the *LCA(i,j)* node itself needs to be (point-)queried or (point-)updated by calling *PointQueryOrUpdate(LCA(i,j))*:

**PointQueryOrUpdate(x):**
**if** (*x* is a super-node) **then** {
 *query or update the node x, taking care to consider the fact that it belongs to node x's group*
} **else if** (*x* belongs to node *y*'s group) **then** {
 *query or update the node x, taking care to consider the fact that it belongs to node y's group*
} **else** {
 *query or update the node x individually, as in the naive solution (which would simply visit every node on the path); node x does not belong to any group*
}

Every time we perform a point query or update on a node belonging to a group we will use the functions for point query and update defined in the block partitioning framework from [1], except that the query aggregate is recomputed only after all the affected nodes of each group were point-updated (just as when updating a partial block). Every time we query or update a full block we will use the corresponding functions from the same framework. The only extra case here occurs when we handle nodes which do not belong to any group (in the framework defined in [1] every cell belonged to some group). These extra cases can be easily handled, by performing individual queries or updates on these nodes (like in a naive solution which would simply traverse all the nodes along the path between the nodes *i* and *j*).

Let's analyze now the time complexity of a path query or update. The number of extra nodes is $O(H)$, the number of nodes which are point-queried or point-updated and belong to a group is also $O(H)$ and the number of full blocks which are queried/updated is $O(N/H)$. Thus, the time complexity is $O(H+N/H)$, the same as in [1]. Basically, by employing the compressed super-node tree representation we were able to extend the results from [1], which were only applicable to a path network, to a tree network without losing any efficiency.

**3.2. Solving a coloring point- and range- query/update problem**

In this sub-section we will define the following problem, which can be handled in the context of the block partitioning framework extension presented in the previous sub-section, but not necessarily at a $O(1)$ simple query/update cost. Each node *i* of the tree has a color *color(i)* and a value *value(i)*. A query *Q(i,j,c)* asks for the aggregate of all the values of the nodes with color *c* on the path between the nodes *i* and *j*. The aggregation function *aggf* is inversible. An update *U(i,j,a,b)* changes the color of all the nodes with color *a* on the path between *i* and *j* to color *b*.

Handling nodes which do not belong to any group is easy. At query time we simply check if their color is *c* and, if so, we "add" their value to the result. During an update, if such a node is colored with color *a*, we simply change its color to color *b*.

For each group *g* we will perform a preprocessing, as follows. Let *SC* be the set of colors existing in the group. We will assign to each color *c* from *SC* a unique identifier *cid(g,c)* (e.g. consecutive numbers from *1* to *|SC|*). *cid* will be

implemented as a hashtable (the key is the color and the value is the identifier). We will also maintain a reverse hashtable for this: *revcid(g,c)*=the color whose unique identifier in group *g* is *c*. Then, for each node *i* in the group, we will maintain a value *colorcid(i)*=the identifier of the color of node *i*. The color identifiers will form a tree structure – each of them will have a parent: *cidparent*; initially, all of these parents are set to an invalid value (e.g. *0*). During the preprocessing stage we will also maintain another hashtable *aggv(g,c)*=the aggregate of the values of all the nodes from the group colored in color *c*.

During an update we will proceed as follows. Let's assume that a node *p* from the group *g* is (point-)updated. First, we will determine the color identifier of its color. In order to do this we will start at *colorcid(p)* and we will follow the *cidparent* pointers until we reach an identifier which has no parent (i.e. it is a *root*). Let this identifier by *ccidx*. Then, the current color of node *p* is *revcid(g,ccidx)*. While following the parent pointers we may also use the well-known path compression heuristic from the disjoint sets data structures [9] in order to reduce the height of the color identifiers tree. What this means is that the *cidparent* pointers of all the color identifiers on the path from *colorcid(p)* (inclusive) to *ccidx* (exclusive) will be set to *ccidx*.

If the current color of node *p* is *x*, then we continue. We will check if the color *y* exists as a key in *cid(g)*. If it doesn't, then we generate a new unique identifier *ccidy* and assign *cid(g,y)=ccidy* and *revcid(g,ccidy)=y*; if it does, then let *ccidy=cid(g,y)*. Then we will set *colorcid(p)=ccidy*. Afterwards, we "subtract" (using $aggf^1$) *value(i)* from *aggv(g,x)* and "add" (using *aggf*) *value(i)* to *aggv(g,y)*. If *x* or *y* do not exist as keys in *aggv(g)* they will first be added with the associated value being the neutral element of the *aggf* operation. If, after the "subtraction", *aggv(g,x)* becomes equal to the neutral element, then *x* may be removed as a key from *aggv(g)*.

Let's assume now that a full group *g* is updated. In this case we will first check if *x* exists as a key in *cid(g)*. If it does, then we continue. Let *ccidx=cid(g,x)*. Then we will check if the color *y* exists as a key in *cid(g)*. If it doesn't, then we generate a new unique identifier *ccidy* and assign *cid(g,y)=ccidy* and *revcid(g,ccidy)=y*; if it does, then let *ccidy=cid(g,y)*. We will set *cidparent(ccidx)=ccidy* and then we will remove *ccidx* as a key from *cid(g)*. *cid(g)* will mainly contain colors in which some nodes from the group are colored; since the update changes all the nodes from the group from color *x* to color *y*, then no node will still be colored in color *x* after the update. Moreover, we will "add" to *aggv(g,y)* the value *aggv(g,x)* (if one or both of the keys *x* and *y* do not exist in *aggv*, then they are first added with the associated value being the neutral element of the aggregation function). After this, *x* is removed as a key from *aggv(g)*. If *aggv(g,y)* is equal to the neutral element of the aggregation function then the key *y* may also be removed from *aggv(g)*.

Let's see now how queries are performed. When a node *p* from a group *g* is point-queried, we will first determine its current color, as shown above. If its color is *c* (the query color) then we will add *value(p)* to the result. When a full group *g* is queried we will simply "add" *aggv(g,c)* to the result (if *c* exists as a key in *aggv(g)*).

The complexity of an update or query operation depends now on the height of the color identifier tree and on the efficiency of the hash table implementation. The hash table may be implemented in order to support $O(1)$ expected time operations [10], but the time complexity of handling the color identifiers seems to be $O(log(H))$ at best (when using the path compression heuristic). Thus, we need to multiply the $O(H+N/H)$ time complexity by the $O(log(H))$ factor when computing the overall time complexity.

Note that this pair of query and update operations can also be supported by the original block partitioning framework presented in [1], if we extend its assumptions a bit. There, it was assumed that query and update aggregates are $O(1)$ in size. In this case, the *aggv* hash table stands for the query aggregate and the color identifier tree plus the *cid* and *revcid* mappings stand for the update aggregate. All of their sizes are $O(H)$ instead of $O(1)$.

### 3.3. Answering a special type of color queries

Let's assume that each node *x* has a color *color(x)*. We need to answer queries of the form *Q(a,b)*=the number of pairs of nodes *(i,j)* such that *color(i)=a*, *color(j)=b* and *i* is an ancestor of *j*.

We will start with the following preprocessing corresponding to nodes which do not belong to any group. We will maintain a global hash table *H*, in which we will have *H(a,b)=c*, where *c* is the number of pairs of nodes *(i,j)* such that *color(i)=a*, *color(j)=b*, *j* is a node belonging to no group and *i* is an acestor of *j* below the lowest super-node *y* located above *j* (excluding *y*). We will consider each node *x* which does not belong to any group and then we will follow the parent pointers starting from *x* until we reach the first super-node *y*. For each visited node *z* (excluding *y*), we will increment *H(color(z),color(x))* by *1* (if the key *(color(z),color(x))* does not exist in the hash table, it is first added with the associated value *0*). Then, for each super-node *y* we will maintain a hash-table *cntA(y,c)*=the number of nodes with color *c* which do not belong to any group and *y* is their lowest super-node ancestor. These values are also computed during the initial step. Once we reach the node *y* starting from a node *x* not belonging to any group we will increment *cntA(y,color(x))* by *1* (if *color(x)* does not exist in *cntA(y)* then we will add it first with the associated value *0*).

The second preprocessing step consists of considering each super-node *x*. We will traverse the tree path starting from *x* and ending at *super-parent(x)* (including *x* and excluding *super-parent(x)*). During the traversal we will maintain a hashtable *cntB(x)*, where *cntB(x,c)*=the number of nodes of color *c* that were traversed so far. Let's assume that we are currently at the node *y* during the path traversal. We will traverse the keys of the hash table *cntB(x)*. For each such key *c*, we will increment *H(color(y),c)* by *cntB(x,c)*. Afterwards we increment *cntB(x,color(y))* by *1*. As before, whenever a key is missing from a hashtable, we add it first with the associated value *0*.

The first two steps take $O(N \cdot H)$ time and memory. After the initial preprocessing we are ready to answer queries. Each query *Q(a,b)* will be answered by traversing the compressed super-node tree bottom-up (from the leaves towards the root). We will maintain a variable *result*, storing the result of the query, which is initialized to *H(a,b)*. Then, during the traversal of the compressed super-node tree, we will compute for each super-node *x* a value *num(x)*=the number of nodes colored with *b* located below *x* (excluding *x*) in the (original) tree. When we reach a super-node *x*, *num(x)* will be equal to the sum of the following values (if any of them do not exist, they will be considered to be *0*):

- *cntA(x,b)*
- the sum of all the values *(num(y,b)+cntB(y,b))*, where *y* is a super-son of *x* (i.e. *y* is a super-node and *x* is *y*'s super-parent)

The value *num(x)* can be computed in time proportional to the number of super-sons that $x$ has (if we maintain the list of $x$'s super-sons after the compressed super-node tree is constructed). Once $num(x)$ is computed, *result* is incremented by $cntB(x,a) \cdot num(x)$.

A query is answered in time proportional to the number of nodes in the compressed super-node tree ($O(N/H)$), but the preprocessing stage takes $O(N \cdot H)$ time and memory. By choosing $H=O(sqrt(N))$ we obtain a preprocessing stage with $O(N \cdot sqrt(N))$ time and memory and a query answering time of $O(sqrt(N))$.

## 4. AMORTIZED HEIGHT BALANCING OF A BINARY SEARCH TREE USING GLOBAL REBUILDING

Let's consider a standard binary search tree in which elements may be inserted and from which elements may be removed. There are many techniques proposed for maintaining a binary search tree balanced [3, 4, 6]. In this section we present a very simple solution, which guarantees that the amortized "cost" of each operation is $O(H+N/H)$, where $N$ is the number of nodes in the tree and $H$ is a parameter (generally a function of $N$).

We will implement the insertions and deletions like in a normal binary search tree (without balancing rules). However, we will maintain in each node of the binary search tree the height of the sub-tree rooted at that node and we will also maintain a global variable $N$ representing the number of nodes in the tree. The root node will contain the height of the entire tree.

Maintaining the height at each node is quite easy. During insertion, the height of the newly inserted node is *0*. Then, while returning from the recursive call (we assume that the insertion was implemented recursively), we recomputed the heights of the visited nodes as *1 + max{height(left son), height(right son) }* (if one of the sons does not exist, its height will be considered *-1*). After a successful insertion, $N$ is incremented by *1*.

A deletion of a node also follows a path in the binary search tree from the root down to a node which is physically removed from the tree. As before, when returning from the recursive calls, the *height* field of the visited nodes will be updated. The variable $N$ will be decremented by *1*. In case logical deletion is used (the node is only marked as deleted but not physically deleted from the tree), no heights will be updated, but $N$ will still be decremented.

General methods for augmenting a binary search tree with extra fields in the nodes were presented in [11].

If, after an insertion or a deletion, the height of the tree exceeds $H$, then we will rebuild the whole tree in a fully balanced manner.

Rebuilding the tree in a balanced manner consists of two steps. First, an in-order traversal of the tree is performed in order to gather all the $N$ elements of the tree in a vector or a list (if we use logical deletion then we ignore the logically deleted elements during the rebuilding phase; thus, during this phase, these elements will, in fact, be discarded and, thus, physically deleted). Then, a recursive procedure is called for rebuilding the tree. Let's say that we need to construct a perfectly balanced sub-tree containing the elements from positions *i* to *j* in the vector/list. The root *r* of the tree will contain the element located on the position $m=(i+j)/2$. Then, if $m>i$ we will call our recursive function for *(i,m-1)* and the root of that sub-tree will be the left son of *r*. If $m<j$ we will call our recursive function for *(m+1,j)* and the root of that sub-tree will be the right son of *r*. Other possibilities for constructing a balanced binary search tree from a sequence of sorted values are discussed in [2].

Let's analyze the time complexity of this approach in the worst case. Each operation (insertion, deletion, search) takes time proportional to the height of the tree. Since in our case the height never exceeds $H$, the time complexity of these operations is $O(H)$. A global rebuilding operation takes $O(N)$ time. Let's assume that we have a sequence of $M$ insertion, deletion and search operations. Each operation takes $O(H)$ time and, in the worst case, every $H$-$log_2(N)$ operations we may need to perform a global rebuilding of the tree. Thus, the total time complexity of the entire sequence of operations is $O(H \cdot M+N \cdot M/(H-log_2(N)))$. Unless $H$ is very close to $log_2(N)$, we can drop the $log_2(N)$ term (when $H$ is already at least $2 \cdot log_2(N)$ the term can be dropped). Thus, we obtain a time complexity of $O(H \cdot M+N/H \cdot M)=O((H+N/H) \cdot M)$. In order to minimize this expression we need to choose $H=O(sqrt(N))$.

## 5. THE H-LIST DATA STRUCTURE

Let's consider an augmented list data structure (which we will call H-list) in which every element of the list has pointers towards its immediate (at most two neighbors), as well as pointers towards the elements located $H$ positions away from it in both directions (we call these nodes H-neighbors). If some neighbor or H-neighbor is missing, the corresponding pointer will be *null*. We want to support two types of queries and two types of updates on these data structures:
- find the endpoint elements, given a pointer to some element $x$ from an H-list
- find the (at most two) elements located $D$ positions away from a given element $x$
- given two elements $x$ and $y$ such that they are both endpoints of two H-lists, concatenate the two H-lists by making the two elements immediate neighbors of each other
- given two elements $x$ and $y$ which are neighbors in an H-list, split the H-list by removing the link between $x$ and $y$

For each element $x$ we will denote the pointers to its two direct neighbors as $v(x,0)$ and $v(x,1)$ and the pointers to its two H-neighbors as $hv(x,0)$ and $hv(x,1)$. Note that the "left" and "right" directions are not properly defined, because of the need to support the concatenate and split operations. However, $hv(x,i)$ will always be the H-neighbor in the same direction as $v(x,i)$.

In order to support all the defined operations we will need the following functions: *opposite_neighbor(x,y)* which returns the neighbor of $x$ different from $y$ and *opposite_hneighbor(x,y)* which returns the H-neighbor of $x$ different from $y$. These functions can be easily implemented: *opposite_neighbor(x,y)=(if y=v(x,1) then v(x,0) else v(x,1))*. Similarly, *opposite_hneighbor(x,y)=(if y=hv(x,1) then hv(x,0) else hv(x,1))*. We will also need the extra functions *neighbor_index(x,y)=(if y=v(x,0) then 0 else 1)* and *hneighbor_index(x,y)=(if y=hv(x,0) then 0 else 1)*.

Let's consider the first query operation, that of finding the endpoints of the H-list, given an element $x$ of the list. We will show how to find the endpoint in direction $i$ away from $x$ and the same algorithm will be used for $i=0$ and $i=1$. If $v(x,i)=null$ then $x$ is an endpoint in the direction $i$. Otherwise, let $y=x$ and $dir=i$. While $hv(y,dir) \neq null$ do: *(1) newy=hv(y,dir); (2) dir=1-hneighbor_index(newy,y); (3) y=newy*. Then, while $v(y,dir) \neq null$ do: *(1) newy=v(y,dir); (2) dir=1-neighbor_index(newy,y); (3) y=newy*. At the end, $y$ is the endpoint when starting from *x* and advancing in direction *i* (relative to *x*).

In order to answer the second type of queries we will proceed similarly. We will consider the direction *i* relative to *x* (*i=0,1*). We will also maintain a counter *cnt* (initially, *cnt=0*). As before, we initialize *y=x* and *dir=i*. Then, the condition of the first while loop becomes: *while (hv(y,dir)≠null and cnt+H≤D)*. Inside the loop we also add a step *(4): cnt=cnt+H*. Then, the condition of the second while loop becomes: *while (v(y,dir)≠null and cnt+1≤D)*. We also add a step *(4)* inside the second while loop: *cnt=cnt+1*. At the end, if *cnt=D* then *y* is one of the elements located *D* positions away from *x* (if *cnt<D* then there is no element located *D* positions away from *x* in the chosen direction).

Assuming that the H-list has *N* elements, each of the first two queries is answered in $O(H+N/H)$ time. In this case, the (maximum) number of elements in the H-list needs to be estimated in advance, as *H* needs to be fixed no matter how many elements actually exist in the H-list.

Let's consider now the concatenation operation. First, we find the values *i* and *j* such that *v(x,i)=null* and *v(y,j)=null*. We will set *v(x,i)=y* and *v(y,j)=x*. Afterwards, we construct two vectors *L(x)* and *L(y)* with the first (at most) *H* nodes of each H-list (starting from *x* and from *y*). We will show how to compute *L(x)* (*L(y)* will be computed similarly). We add the pair *(x,i)* as the first element of *L(x)* and then we initialize *a=x, b=opposite_neighbor(x,y)* and *cnt=1*. While (*b≠null* and *cnt<H*) do: *(1)* add the pair *(b, neighbor_index(b,a))* at the end of *L(x); (2) c=opposite_neighbor(b,a); (3) a=b; (4) b=c; (5) cnt=cnt+1*.

After constructing both *L(x)* and *L(y)* (we consider them indexed from *1*), we proceed as follows. Let *(a,b)* be the pair located on position *c* in *L(x)* and *(d,e)* the pair on position *H-c+1* (we will start with *c=*the last position of *L(x)* and move towards *c=1*; we will stop after handling the case *c=1* or when *H-c+1* exceeds the size of *L(y)*). We will set *hv(a,b)=d* and *hv(d,e)=a*.

When performing a split between two elements *x* and *y* we will start by finding the directions *i* and *j* such that *v(x,i)=y* and *v(y,j)=x*. Then we will compute the lists *L(x)* and *L(y)* as before. For each pair *(a,b)* from *L(x)* or *L(y)* we will set *hv(a,b)=null*. In the end we will set *v(x,i)=v(y,j)=null* (i.e. we break the link between *x* and *y*).

Both types of update operations (concatenation and split) take $O(H)$ time.

## 6. RELATED WORK

Heavy-path decompositions and longest-path decompositions [7, 8] are generic techniques which also construct a compressed tree representation of a general rooted tree. Each node of the compressed tree corresponds to a path in the tree and each edge corresponds to an edge of the original tree. Using these techniques, any problem that can be solved for paths (sequences) can also be solved for trees with an $O(log(N))$ (for the heavy-path decomposition) or $O(sqrt(N))$ (for the longest-path decomposition) extra factor added to the time complexity. The block partitioning framework from [1] can easily be extended to these decompositions by constructing a block partitioning for each path from a node of the compressed tree. In this case the extra factor added to the time complexity is probably somewhat better – since the worst case seems to occur when we encounter $O(log(N))$ paths each having $O(N/log(N))$ nodes, the time complexity in this case is $O(log(N) \cdot sqrt(N/log(N)))=O(sqrt(log(N)) \cdot sqrt(N))$. However, the compressed super-node tree representation introduced in this paper does not introduce any extra time complexity overhead. Moreover, it seems to be applicable to a somewhat wider range of problems than the heavy-path and longest-path tree decompositions (because both the height and the number of nodes of our compressed super-node tree are small), however at the expense of a potentially more time and memory consuming preprocessing step (because large „chunks" of the tree are actually „outside" of the compressed super-node tree representation, as nodes not belonging to any group).

Techniques for maintaining balanced binary search trees were discussed in many papers and books [2, 3, 4, 6], including tree rebuilding techniques (only partial rebuilsing in some cases).While the global rebuilding method presented in this paper is quite inefficient in terms of performance guarantees compared to the more sofisticated methods proposed in the literature, it is, nevertheless, quite easy to implement.

Efficient list-based data structures were proposed in [5]. Skiplists [5] support both concatenation and splitting, together with many types of queries (although some of the queries we considered seem to only be supported in one direction). As before, we bring the simplicity argument in favor of our H-list data structure. Other data structures which may support the same operations as the H-list are some classes of balanced binary search trees [6].

## 7. CONCLUSIONS AND FUTURE WORK

In this paper we presented several novel $O(H(N)+N/H(N))$ algorithmic techniques for multiple query and update problems on rooted trees and lists. The compressed super-node tree representation of a general rooted tree is a data structure which can be used for solving similar problems to other standard tree decompositions (e.g. the heavy-path or longest-path tree decompositions), but at a better time complexity in some situations. The H-list and the global rebuilding technique proposed for binary search trees are less efficient than other solutions proposed in the scientific literature, but have the advantage that they are quite easy to implement. As future work we intend to perform an experimental evaluation of the techniques proposed in this paper and compare them with other methods proposed in the literature on a meaningful set of query and update operations.

## 8. ACKNOWLEDGEMENTS

The work presented in this paper was funded by CNCS-UEFISCDI under research grant PD_240/2010 (AATOMMS - contract no. 33/28.07.2010), and by the Sectoral Operational Programme Human Resources Development 2007-2013 of the Romanian Ministry of Labour, Family and Social Protection through the financial agreement POSDRU/89/1.5/S/62557.

## 9. REFERENCES

[1] **M. I. Andreica:** *Optimal Scheduling of File Transfers with Divisible Sizes on Multiple Disjoint Paths*, Proceedings of the 7[th] IEEE Romania International Conference "Communications", pp. 155-158, 2008.
[2] **J. G. Vaucher:** *Building Optimal Binary Search Trees from Sorted Values in O(N) Time*, Essays in Memory of Ole-Johan Dahl, vol. 2635, pp. 376-388, Springer, 2004.
[3] **D. Samanta:** *Classic Data Structures, 2[nd] Edition*, 2009.

**[4] I. Galperin, R. L. Rivest:** *Scapegoat Trees*, Proceedings of the 4[th] Annual ACM-SIAM Symposium on Discrete Algorithms, pp. 165-174, 1993.
**[5] W. Pugh:** *Skip Lists: A Probabilistic Alternative to Balanced Trees*, Communications of the ACM, vol. 33 (6), pp. 668-676, 1990.
**[6] C. R. Aragon, R. Seidel:** *Randomized Search Trees*, Proceedings of the 30[th] Symposium on Foundations of Computer Science, pp. 540-545, 1989.
**[7] M. I. Andreica, E.-D. Tirsa:** *Towards a Real-Time Scheduling Framework for Data Transfers in Tree Networks*, Proceedings of the 10[th] IEEE International Symposium on Symbolic and Numeric Algorithms for Scientific Computing (SYNASC), pp. 467-474, 2008.
**[8] D. Harel, R. E. Tarjan:** *Fast Algorithms for Finding Nearest Common Ancestors*, SIAM J. Comput, vol. 13, pp. 338-355, 1984.
**[9] D. C. Kozen:** *The Design and Analysis of Algoithms*, Springer, 1992.
**[10] E. B. Koffman, P. A. T. Wolfgang:** *Data Structures: Abstraction and Design using Java*, John Wiley & Sons, 2010.
**[11] T. H. Cormen, C. E. Leiserson, R. L. Rivest, C. Stein:** *Introduction to Algorithms, 2[nd] Edition*, MIT Press, 2001.
**[12] M. E. Andreica, I. Dobre, M. Andreica, B. Nitu, R. Andreica:** *A New Approach of the Risk Project from Managerial Perspective*, Economic Computation and Economic Cybernetics Studies and Research, no. 1-2, pp. 121-130, 2008.
**[13] S. Farmache, M. Andreica:** *Improving Stock Management Assessment Instruments*, Metalurgia International, vol. 15, no. 3, pp. 167-170, 2010.